\documentclass[preprint,aps,prd,superscriptaddress,nofootinbib]{revtex4}%
\usepackage{graphicx}
\usepackage{amsmath}
\usepackage{amssymb}
\usepackage{bm}
\usepackage{epsfig}
\usepackage{ulem}

\begin{document}
\preprint{}
\title{\mbox{}\\[10pt]
Comment on\\ ``Single-inclusive jet production in electron-nucleon 
collisions\\ through next-to-next-to-leading order in perturbative QCD''\\
{[Phys.\ Lett.\ B763, 52--59 (2016)]}}
\author{Geoffrey~T.~Bodwin}
\email[]{gtb@anl.gov}
\affiliation{High Energy Physics Division, Argonne National Laboratory,
Argonne, IL 60439, USA}
\author{Eric Braaten}
\email[]{braaten@mps.ohio-state.edu}
\affiliation{Department of Physics, The Ohio State University, Columbus, 
Ohio 43210, USA}

\date{\today}
\begin{abstract}
In the cross section for single-inclusive jet production in
electron-nucleon collisions, the distribution of a quark in an electron
appears at next-to-next-to-leading order. The numerical calculations in
Ref.~\cite{Abelof:2016pby} were carried out using a perturbative
approximation for the distribution of a quark in an electron. We point
out that that distribution receives nonperturbative QCD contributions
that invalidate the perturbative approximation. Those nonperturbative
effects enter into cross sections for hard-scattering processes through
resolved-electron contributions and can be taken into account by
determining the distribution of a quark in an electron
phenomenologically.
\end{abstract}

\vspace{-1cm}
\maketitle


In Ref.~\cite{Abelof:2016pby}, the cross section for single-jet
inclusive production in lepton-nucleon collisions is computed through
next-to-next-to-leading order in perturbative quantum chromodynamics
(QCD). That computation advances significantly the potential for
precision comparisons between theory and experiment for this process.
The cross section contains a contribution that is proportional to the
distribution of a quark in a lepton, namely, $f_{q/l}(\xi,\mu^2)$, where
$\xi$ is the light-cone momentum fraction of the quark and $\mu$ is the
renormalization scale. Such a contribution could be termed a
``resolved-lepton'' contribution. The distribution that was used in
Ref.~\cite{Abelof:2016pby} is
\begin{eqnarray}
f_{q/l}(\xi,\mu^2) &=& e_q^2 \left(\frac{\alpha}{2 \pi} \right)^2 \left\{
\left[ \frac{(1-\xi)(4+7\xi+4\xi^2)}{6\xi} +(1+\xi)\log\xi \right] \log^2\frac{\mu^2}{m_l^2} \right.
\nonumber\\
&& \hspace{-2cm} \left.
+ \left[- \frac{(1-\xi)(2+5\xi-2\xi^2)}{\xi} - \frac{8+15\xi-3\xi^2-8\xi^3}{3\xi} \log\xi
-3(1+\xi)\log^2\xi \right] \log\frac{\mu^2}{m_l^2} \right\},
\label{fq/l}
\end{eqnarray}
where $m_l$ is the lepton mass, $e_q$ is the electric charge of the
quark, and $\alpha$ is the quantum-electrodynamics (QED) coupling
constant. The single and double logarithms of $\mu$ cancel the
$\mu$-dependence of other factors in the cross section at order
$\alpha^2\alpha_s^2$.

In Ref.~\cite{Abelof:2016pby},
$f_{q/l}(\xi,\mu^2)$ is derived by making use of the Dokshitzer, Gribov,
Lipatov, Altarelli, Parisi (DGLAP) evolution equation
\cite{Gribov:1972ri, Lipatov:1974qm, Dokshitzer:1977sg, Altarelli:1977zs}
in the form
\begin{equation}
\mu^2\frac{\partial ~~}{\partial\mu^2} f_{q/l} =
P_{q\gamma}\otimes f_{\gamma/l} + P_{ql}\otimes f_{l/l}.
\label{DGLAP-orig}
\end{equation}
Here, $f_{\gamma/l}(\xi,\mu^2)$ is the distribution of a photon in a
lepton, $f_{l/l}(\xi,\mu^2)$ is the distribution of a lepton in a
lepton, $P_{q\gamma}(z)$ and $P_{ql}(z)$ are the DGLAP splitting
functions, and $\otimes$ denotes the convolution
\begin{equation}
[P\otimes f](\xi)=\int_\xi^1\frac{dz}{z} P(\xi)f(\xi/z).
\end{equation}
(In Eq.~(\ref{DGLAP-orig}), we have absorbed factors of $\alpha$ into the
definitions of the splitting functions.) In Ref.~\cite{Abelof:2016pby},
the splitting functions are evaluated to order $\alpha$ and order
$\alpha^2$, respectively, and the QED distributions on the right side of
Eq.~(\ref{DGLAP-orig}) are evaluated at leading order in $\alpha$:
$f_{\gamma/l}(\xi,\mu^2)$ is the Weizs\"acker-Williams distribution, and
$f_{l/l}(\xi) = \delta(1-\xi)$. The distribution in Eq.~\eqref{fq/l} is
obtained by integrating  Eq.~(\ref{DGLAP-orig}) with the boundary condition
$f_{q/l}(\xi,m_l^2)=0$.

In this comment, we  point out that $f_{q/l}(\xi,\mu^2)$ receives
nonperturbative QCD contributions that invalidate the expression for the
distribution of a quark in an electron defined by Eq.~\eqref{fq/l}. If
the lepton has a sufficiently large mass, as is the case for the 
$\tau$ lepton, then $f_{q/l}(\xi,m_l^2)$ can
be computed in QCD perturbation theory, and it can be evolved
perturbatively from the scale $m_l^2$ to the scale $\mu^2$ in order to
absorb logarithms of $\mu^2/m_l^2$ into $f_{q/l}(\xi,\mu^2)$. In this
case, the expression in Eq.~\eqref{fq/l} is a valid approximation for
$f_{q/l}(\xi,\mu^2)$ in that it captures the logarithmic contributions
at leading-order in $\alpha$.\footnote{We note that the expression in
Eq.~\eqref{fq/l} omits constant terms that arise in standard
renormalization schemes, such as modified minimal subtraction.} However,
when the lepton is an electron or a muon, $f_{q/l}(\xi,\mu^2)$ cannot be
computed in QCD perturbation theory.

The nonperturbative nature of $f_{q/l}(\xi,\mu^2)$ can be seen by
considering its DGLAP evolution. When one considers QCD corrections, the
evolution equation for $f_{q/l}(\xi,\mu^2)$ contains additional
contributions that arise from the emission of real and virtual gluons by
the quark:
\begin{eqnarray}
&&\mu^2\frac{\partial ~~}{\partial \mu^2}               
\begin{pmatrix} f_{q_i/l} \\ f_g \end{pmatrix}
=\begin{pmatrix} P_{q_i\gamma} \otimes f_{\gamma/l} \\ 0 \end{pmatrix}
+\begin{pmatrix} P_{q_il}\otimes f_{l/l} \\ 0 \end{pmatrix}
 +\sum_{q_j}
\begin{pmatrix} P_{q_iq_j} & 2 P_{q_ig} \\ P_{gq_j} & P_{gg} \end{pmatrix}
\otimes 
\begin{pmatrix} f_{q_j/l} \\ f_{g/l} \end{pmatrix},
\label{DGLAP-full}
\end{eqnarray}
where the sum over $q_j$ includes both quarks and antiquarks. Suppose
that one were to follow the procedure in Ref.~\cite{Abelof:2016pby},
evolving $f_{q/l}$ from the scale $m_l$ to a hard-scattering scale. The
splitting functions in Eq.~(4) depend on $\alpha_s$ at scales $\mu$ that
range from $m_l$ to the hard-scattering scale. If $\mu$ is sufficiently
large, then the splitting functions can be computed in perturbation
theory. However, if $\mu$ is less than a scale of order
$\Lambda_\mathrm{QCD}$, then the perturbation expansion for the
splitting functions fails, and the evolution of $f_{q/l}$ receives
nonperturbative contributions. In the case of the electron or the muon,
the range of $\mu$ includes a region in which perturbative QCD fails and
nonperturbative effects dominate.

Although the computation of the short-distance part of the cross section
through the order of interest in Ref.~\cite{Abelof:2016pby} requires
only that collinear poles through order $\alpha^2$ be absorbed into
$f_{q/l}(\xi,\mu^2)$, a reliable calculation of $f_{q/l}(\xi,\mu^2)$
requires that QCD corrections be taken into account. The concept that the
short-distance part of the cross section can be computed at a fixed
order in $\alpha_s$, while the parton distributions, when they are
nonperturbative, cannot is, of course, familiar from other
hard-scattering processes, such as deep-inelastic scattering.

The nonperturbative distribution for a quark in an electron
$f_{q/e}(\xi,\mu^2)$ at a scale $\mu^2$ that is in the perturbative
regime of QCD could, in principle,  be determined phenomenologically by
fitting cross-section predictions to data. A process that is particularly
sensitive to $f_{q/e}(\xi,\mu^2)$ is single-inclusive jet production in
electron-electron scattering. Alternatively, with some sacrifice of
sensitivity, one could make use of cross sections for single-jet
inclusive production in electron-nucleon collisions. Lattice
calculations might also provide information on $f_{q/e}(\xi,\mu^2)$.
Once the nonperturbative distribution for a quark in an electron has
been determined, it could be used to make reliable predictions for the
resolved-electron contributions to hard-scattering processes.

Because of the sensitivity of $f_{q/e}(\xi,\mu^2)$ to nonperturbative
QCD effects, the expression in Eq.~\eqref{fq/l} can at best be regarded
as a model for the distribution. One unphysical aspect of this model is
its double-logarithmic dependence on the electron mass. There is a
logarithm of $m_e^2$ in the Weizs\"acker-Williams distribution
$f_{\gamma/e}(\xi,\mu^2)$.  A second logarithm arises when one
integrates Eq.~\eqref{DGLAP-orig} from $m_e^2$ to $\mu^2$ using the 
perturbative expressions for the splitting functions. This
procedure implies that quarks in the electron are generated by
perturbative evolution all the way down to virtualities of order
$m_e^2$.  One would not expect a probe with a virtuality that is much
less than a typical hadronic scale to be able to resolve the hadronic
structure of the electron. For the range of $\mu$ that is considered in
Ref.~\cite{Abelof:2016pby}, much of the large coefficient
$\log^2(\mu^2/m_e^2)$ in Eq.~\eqref{fq/l} comes from integration over
virtualities that are smaller than a typical hadronic scale of, say,
700~MeV. This feature of the model in Eq.~\eqref{fq/l} would tend to
produce a significant overestimate of the contribution from quarks in
the electron to the cross section for single-jet inclusive production in
electron-nucleon collisions. Other nonperturbative effects that are not
accounted for in the model could be substantial, as well.

We note that a sensitivity to nonperturbative QCD effects arises in
the same way in the case of the distribution of a quark in a real photon
$f_{q/\gamma}$. In this case, the leading-order QED expression for the
logarithmic contribution to the distribution that is analogous to
Eq.~\eqref{fq/l} is
\begin{equation}
f_{q/\gamma}(\xi,\mu^2)=e_q^2\frac{\alpha}{2\pi}[\xi^2+(1-\xi)^2]
\log \frac{\mu^2}{m_\gamma^2}.
\label{f-q-gamma}
\end{equation}
The inadequacy of this leading-order logarithmic approximation  
is manifest in the
logarithm of the photon mass $m_\gamma$. Of course, it is well
established that the distribution of a quark in a real photon involves
contributions that cannot be calculated in perturbation theory, but
must, instead, be obtained from fits to experimental data. (See, for
example, Refs.~\cite{Cornet:2004nb,Aurenche:2005da,Berger:2014rva}.)

\begin{acknowledgments}

We thank Frank Petriello and Yuri Kovchegov for helpful discussions.
The work of E.B.\ was supported in part by the Department of Energy
under grant DE-SC0011726. The work of G.T.B.\ is supported by the U.S.\
Department of Energy, Division of High Energy Physics, under Contract
No.\ DE-AC02-06CH11357. The submitted manuscript has been created in
part by UChicago Argonne, LLC, Operator of Argonne National Laboratory.
Argonne, a U.S.\ Department of Energy Office of Science laboratory, is
operated under Contract No.\ DE-AC02-06CH11357. The U.S. Government
retains for itself, and others acting on its behalf, a paid-up
nonexclusive, irrevocable worldwide license in said article to
reproduce, prepare derivative works, distribute copies to the public,
and perform publicly and display publicly, by or on behalf of the
Government.

\end{acknowledgments}


\end{document}